\pdfoutput=1
\documentclass[twocolumn,letterpaper]{IEEEAerospaceCLS}  

\usepackage[]{graphicx}    
\usepackage{epstopdf}
\usepackage{gensymb, amsmath, color}
\usepackage{hyperref}
\newcommand{\ignore}[1]{}  

\def\skiplinehalf{\medskip\\}

\def\supit#1{\raisebox{0.8ex}{\small\it #1}\hspace{0.05em}}

\begin{document}
\title{A High-resolution Pointing System for Fast Scanning Platforms: the EBEX Example}

\author{%
Joy Didier\supit{a}, Daniel Chapman\supit{a}, Asad M. Aboobaker\supit{b}, Derek Araujo\supit{a}, \\
Will Grainger\supit{c}, Shaul Hanany\supit{d}, Kyle Helson\supit{e}, Seth Hillbrand\supit{h}, \\
Andrei Korotkov\supit{e}, Michele Limon\supit{a}, Amber Miller\supit{a}, Britt Reichborn-Kjennerud\supit{a}, \\
Ilan Sagiv\supit{f}, Greg Tucker\supit{e}, Yuri Vinokurov\supit{g}\\
\skiplinehalf
\supit{a}Columbia University, New York, NY 10027, U.S.A.;\\
\supit{b}Jet Propulsion Laboratory, Pasadena, CA 91101, U.S.A;\\
\supit{c}Rutherford Appleton Lab, Harwell Oxford, OX11 0QX, U.K.;\\
\supit{d}University of Minnesota, Minneapolis, MN 55455, U.S.A.;\\
\supit{e}Brown University, Providence, RI 02912, U.S.A.;\\
\supit{f}Weizmann Institute of Science, Rehovot 76100, Israel;\\
\supit{g}Carnegie Mellon University, Pittsburgh, PA 15213, U.S.A.;\\
\supit{h}California State University, Sacremento, CA 95819, U.S.A.;
\thanks{\footnotesize \copyright2015 IEEE. Personal use of this material is permitted. Permission from IEEE must be obtained for all other uses, in any current or future media, including reprinting/republishing this material for advertising or promotional purposes, creating new collective works, for resale or redistribution to servers or lists, or reuse of any copyrighted component of this work in other works.}
\thanks{\footnotesize \href{http://ieeexplore.ieee.org/xpl/articleDetails.jsp?arnumber=7119010}{DOI 10.1109/AERO.2015.7119010}}
\thanks{\footnotesize 978-1-4799-5380-6/15/$\$31.00$}
}

\maketitle

\thispagestyle{plain}
\pagestyle{plain}

\begin{abstract} 

The E and B experiment (EBEX) is a balloon-borne telescope designed to measure the polarization of the cosmic microwave background with 8' resolution employing a gondola scanning with speeds
of order degree per second. In January 2013, EBEX completed 11 days of observations in a flight over Antarctica covering $\sim$6000 square degrees of the sky. The payload is equipped with two
redundant star cameras and two sets of three orthogonal gyroscopes to reconstruct the telescope attitude.  The EBEX science goals require the pointing to be reconstructed to approximately
10" in the map domain, and in-flight attitude control requires the real time pointing to be accurate to $\sim$0.5\degree.  The high velocity scan strategy of EBEX coupled to its float
altitude only permits the star cameras to take images at scan turnarounds, every $\sim$40 seconds, and thus requires the development of a pointing system with low noise gyroscopes and
carefully controlled systematic errors.  Here we report on the design of the pointing system and on a simulation pipeline developed to understand and minimize the effects of
systematic errors. The performance of the system is evaluated using the 2012/2013 flight data, and we show that we achieve a pointing error with RMS=25" on 40 seconds azimuth throws, corresponding to
an error of $\sim$4.6" in the map domain.
\end{abstract}

\tableofcontents

\section{Introduction}
EBEX \cite{ebex} is a balloon-borne telescope designed to measure the polarization of the Cosmic Microwave Background (CMB). The 6,000 lb instrument collects data while suspended from a
$10^6$ cubic meter stratospheric balloon at an altitude of $\sim$ 35 km that circumnavigates Antarctica. This paper describes the pointing system that was designed to meet the attitude
requirements of the EBEX telescope, both for in-flight and post-flight attitude determination.  It focuses on the steps taken to meet the post-flight requirements given the timescales
between star camera images.  Section \ref{section: the instrument} describes the instrument and the attitude control system. Section \ref{section: pointing reconstruction} describes
the software developed to reconstruct the attitude.  Section \ref{section: systematics} details the systematic errors in the system and their effect on the pointing reconstruction. Section
\ref{section: simulation_pipeline} describes the simulation pipeline developed to fine tune the pointing reconstruction software and to assess how well systematic errors needed to be
constrained. Section \ref{section: flight} describes the 2012/2013 Antarctic flight and section \ref{section: results} evaluates the performance of the reconstruction software both for
simulations and for flight data.

\section{The EBEX Pointing System}
\label{section: the instrument}

\subsection{Pointing Requirements} 

The requirements on the accuracy of the attitude are different for in-flight real-time pointing, and for post-flight reconstructed pointing used to make CMB maps. The real-time
pointing requirement is set by the need to ensure that detectors are scanning the desired part of the sky. For every detector sample, an accuracy of $\sim$ 0.5\degree is required,
such that the error is small compared to the 6\degree \space EBEX focal plane. The requirement on the accuracy of post-flight reconstructed pointing is set by the EBEX science
goals.  The requirement originally exists in the map domain: the average of all pointing errors inside a given pixel of size 2' should be less than approximatively 10".  Therefore
the requirement on the pointing in the time domain is dependent on the scan strategy and the number of repeated hits per pixel.  To generalize our results, we use our known scan
pattern to transform this requirement in terms of a requirement on the error over a 40 second azimuth throw. Details of this transformation can be found in appendix
\ref{appendix: transform requirements}. We define the error on a throw as the RMS of the difference between the true pointing and the reconstructed pointing for the set of
points between two star camera images. For the EBEX Long Duration (LD) flight scan strategy, the RMS on a 40 second throw needs to be smaller than 54". It is also interesting to
compare any result to the error coming from integrating pure gyroscope rate white noise on a 40 second throw, which is 11" (see appendix \ref{appendix: white_noise_rms} for
details).

\subsection{The gondola and the attitude control system}

\begin{figure}
\centering
\includegraphics[width=3.25in]{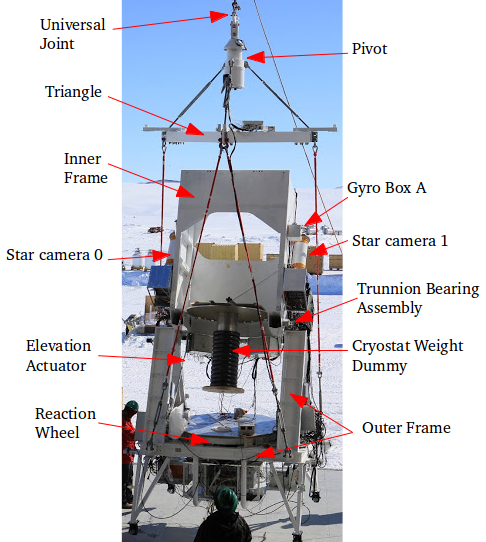}
\caption{The EBEX instrument in Antarctica before the Long Duration flight. The cryostat holding the cold optics and detectors is not mounted on the telescope in the picture, and is
replaced by equivalent weights instead.}
\label{fig: telescope}
\end{figure}

The EBEX instrument, shown in figure \ref{fig: telescope}, is a microwave telescope mounted in an aluminium structure called the gondola. The gondola hangs from an aluminium triangle spreader
bar through four synthetic ropes.  The triangle hangs from a rotator motor (the pivot) and a universal joint, which at float connect to the flight train and balloon.  The azimuth motion is
powered through the pivot motor. A motorized reaction wheel on the gondola provides fine control of the azimuth motion.  The gondola consists of an inner frame and an outer frame connecting
at the trunnion bearing assembly. The inner frame can move in the elevation direction through the elevation actuator. The Attitude Control System (ACS) of the gondola was inherited from
BLAST \cite{blast} and adapted for EBEX. It performs two tasks. Pointing sensors mounted on the gondola collect attitude information which is transmitted to the flight control program (FCP).
FCP uses sensor information to estimate the gondola's attitude, which is then used by the ACS electronics to send currents to the azimuth and elevation motors to orient the telescope
according to a pre-designed scan strategy. The current sent to the motors is governed by PI loops based on position or velocity, and the P and I values need to be tuned in flight to ensure
optimal motion of the telescope. The FCP control loop runs at 100.16 Hz and sensor data is recorded on disk at the same frequency.

\subsection{Star Cameras} 

EBEX has two star cameras, mounted on either side of the inner frame and roughly aligned with the telescope beam, as shown in figure \ref{fig: telescope}. They operate by taking pictures of
the stars and comparing the images to known star catalogs. This process is referred to as finding a pointing solution. Each star camera consists of a telephoto lens, a CCD and a computer
mounted in a rigid assembly inside a pressure vessel, as shown in figure \ref{fig: other sensors}. The computer runs the Star Tracking Attitude Reconstruction Software (STARS) \cite{chappy
spie}, a platform-independent software developed for EBEX in C++ that captures the images, finds the bright spots in the image, matches their pattern to a known catalog of stars and
communicates with FCP. The precision of the star cameras is mainly determined by the CCD and the lens. Each EBEX star camera contains a 200 mm f/1.8 lens\footnote{Canon EF 200 mm F/1.8 L USM
Lens} and a CCD\footnote{Kodak KAF-1603E image sensor} with 1536x1024 9 $\mu m$ pixels. The camera has a FWHM of roughly 9", which is also the approximate size of a pixel.  The atmospheric
brightness at 35km altitude requires the star cameras to take 300 ms exposures in order to see stars.  Given the necessary integration time, images must be taken at velocities under 2'/s to
avoid motion-blur. The pointing solutions have 1.5" accuracy in displacement angle on the sky, and 48" accuracy in rotation around the image center. A detailed review of the star cameras
design and in-flight performance can be found in the Chapman et al. companion paper \cite{chappy ieee paper}. 


\subsection{Gyroscopes}

EBEX has two redundant sets of three orthogonal fiber optic gyroscopes\footnote{KVH-DSP 3000}. Each gyroscope measures the rate of angular rotation around its axis. Each set of
three gyroscopes is mounted inside an aluminium box such that they form a nearly orthogonal frame. Box A is mounted near star camera 1, as shown in figure \ref{fig: telescope}.
Box B, not visible in the figure, is mounted near the secondary mirror on the inner frame. Each gyroscope outputs a digital signal at 1000 Hz which is read out by an on-board
Digital Signal Processing (DSP) unit. The data is de-spiked and passed through a stage 4 box-car IIR filter, and written to disk at 100.16 Hz. The filter introduces a measured 33
ms time lag that is accounted for in the pointing reconstruction. To reconstruct the gondola attitude, the gyroscopes' angular rates are integrated in between star camera readings,
taken approximatively every 40 seconds. 

\subsection{Coarse sensors} 

\begin{figure*}
\centering \includegraphics[scale=0.6]{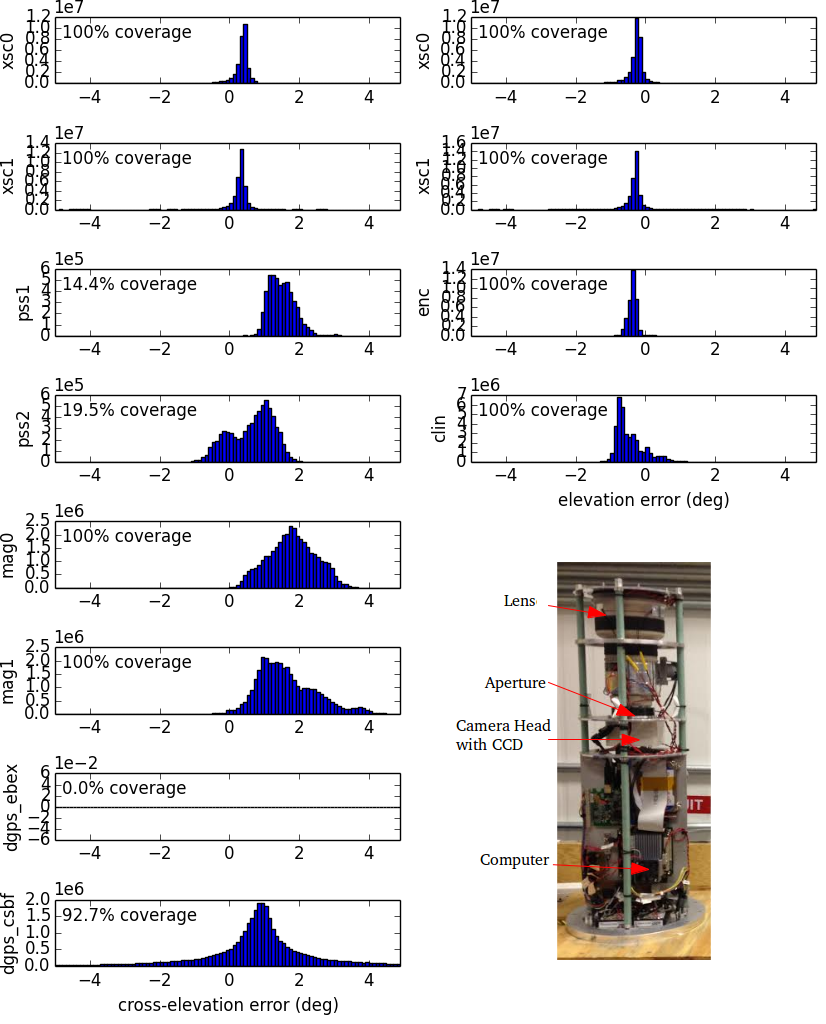} 
\caption{{\bf{Bottom Right:}} One of EBEX star cameras. {\bf{Left and Top Right:}} The performance in cross-elevation (defined as azimuth$\times$cos(elevation), left column) and elevation (right column) of each
absolute pointing sensor during the 2012/2013 EBEX flight.  Each plot shows a histogram of the differences between the post-flight reconstructed boresight pointing, and the in-flight sensor
pointing computed using pre-flight offset calibration. This is a fair representation of how sensors would have performed real time had the star cameras not provided real-time solutions.
During flight, the sensors offsets were re-computed with available star camera solutions. This improved the accuracy of many of the sensors as the pre-flight calibration procedure is limited
by systematic errors specific to ground measurements, such as magnetic fields, atmospheric refraction or signal interference.  Note that not all sensors had valid pointing streams throughout the
flight, as indicated by the percentage covered shown in each plot.  Most notably the sun sensors had very limited coverage because they did not cover the full azimuth range, and the
EBEX differential GPS had no coverage at float.}
\label{fig: other sensors}
\end{figure*}

Balloon-borne instruments need reliable, redundant pointing systems both because minimal intervention is possible once the instrument is at float and also because the temperature,
pressure and radiation environment at float is very different from that on the ground, making end-to-end testing conditions difficult to reproduce before the flight.  For this
reason EBEX was equipped with redundant primary sensors (two star cameras and two sets of orthogonal gyroscopes), but also with an array of less accurate but simpler,  more
reliable sensors: magnetometers, sun sensors, clinometers, two differential GPS, and an elevation encoder. These sensors cannot provide the accuracy required for post-flight
pointing reconstruction, but can provide real-time attitude both to provide a pointing guess to the star camera and in the eventuality that the star cameras cannot solve images in
real-time. The real-time attitude of the gondola is computed in the following manner. FCP keeps a time stream of the attitude of each sensor. For each sensor the attitude in
between new readings is estimated by integrating the gyroscope rates.  When a new sensor reading is available, it is averaged with the integrated attitude time stream.  Then to
compute the gondola's attitude, at every sample in time, a weighted average of functional sensors' attitude is performed.  The main source of error for in-flight pointing is the
proper calibration of each sensor offset to the star cameras (and through the star cameras to the microwave beam). To provide an estimate of these sensor offsets before flight, an
outdoor calibration of all sensors to the star cameras was done in Antarctica. During the test, the star cameras provided solutions using the few stars bright enough to be visible
from the ground during the Antarctic summer. All sensor offsets are re-computed in flight as soon as a star camera solution is available.  Figure \ref{fig: other sensors} lists all
sensors and plots their in-flight accuracy given the calibration performed {\textit{pre-flight}}. The pre-flight calibration is used to show what would be the performance of the
sensors in providing the star camera with a pointing guess and in providing real-time pointing before the star cameras solved. During the EBEX 2012/2013 LD flight, the star cameras
provided pointing solutions in real-time and thus were used as our primary sensors both for real-time and post-flight pointing.  The rest of our analysis will thus focus on star
cameras and gyroscopes only.

\section{Pointing Reconstruction} 
\label{section: pointing reconstruction}

\subsection{Reconstructing the attitude} 

The pointing reconstruction software takes as inputs star camera solutions and gyroscope rates, and outputs the telescope attitude.  The
attitude between two star camera images is reconstructed by integrating the gyroscope angular velocities. The output pointing consists of a time stream where each sample provides
the attitude information to orient the telescope in a given reference frame, chosen here to be the equatorial reference frame.  Quaternions, Euler angles and rotation matrices
provide equivalent frameworks to describe the orientation of the gondola at every time step. In this paper we will use Euler angles, using conventions such that the first rotation
corresponds to the equatorial Right Ascension (RA) angle, the second rotation corresponds to the equatorial Declination (Dec) angle, and the third rotation around the star camera
beam corresponds to Roll. The pointing reconstruction software thus outputs RA, Dec and Roll angles for every time step.  

\subsection{Unscented Kalman Filter} 

We adapted an Unscented Kalman Filter (UKF) \cite{ukf} to C++ to determine the pointing for EBEX given the star camera and gyroscope
measurements. Kalman filters have been extensively described in the literature \cite{kalman} to estimate the state of linear systems given a set of noisy measurements. The
algorithm works in a two step process. In the "prediction" step, the filter produces an estimate of the next attitude and its covariance given the current attitude and by
integrating the gyroscopes, using equation \ref{eq: propagation}:

\begin{equation}
\label{eq: propagation}
\left( 
\begin{array}{c}
\dot{\theta} \\ \dot{\psi} \\ \dot{\phi} 
\end{array} 
\right) = 
\left( 
\begin{array}{ccc} 
c_{\psi}              & 0 & s_{\psi} \\
s_{\psi} t_{\theta} & 1 & -c_{\psi} t_{\theta} \\
-\frac{s_{\psi}}{c_{\theta}} & 0 & \frac{c_{\psi}}{c_{\theta}}
\end{array} 
\right) 
R
O
\left( 
\begin{array}{c} 
\omega_{1} \\ \omega_{2} \\ \omega_{3}
\end{array} 
\right) 
\end{equation}

where $c$, $s$ and $t$ stand for cosine, sine and tangent,  $\theta$, $\psi$, $\phi$ are the attitude Euler angles for Dec, Roll and RA, and $\omega_1$, $\omega_2$, $\omega_3$ are
the gyroscope rates. R and O, called the rotation and the orthogonalization matrix, represent the transformation necessary to rotate each gyroscope into one of the star camera
frame axes. They are detailed in the next subsection.  The variance of the error on the attitude grows as subsequent gyroscope samples are integrated between star camera
measurements.  The system of equations to integrate the Euler angles given the gyroscope rates is highly non-linear, and it is thus necessary to use a generalization of the
Kalman filter in order to propagate the attitude and the covariance. The UKF provides a good balance between Extended Kalman Filters, which linearize the system only to first
order, and full Monte Carlo simulations, which take significantly longer to run. To propagate the attitude and the covariance, the UKF uses a deterministic sampling technique known
as the unscented transform \cite{unscented transform} to pick a minimal set of sampling points around the current attitude.  These points are propagated through the full non linear
equations.  The next attitude and covariance are computed using the mean and covariance of the propagated points.  When a new star camera solution is available, the "update" step
is performed. The estimated attitude is updated using a weighed average of the integrated attitude and the new star camera reading. When the star camera
solutions are good, this weighted solution is dominated by the star camera measurement. The error on the attitude drops, and the process is repeated on the next throw. 

Section \ref{section: systematics} will show that a constant gyroscope rate offset must be fit in order for the RMS of the reconstructed pointing to satisfy the requirements.
Finding this additional parameter is possible because every new star camera reading provides an accurate measurement of the current attitude. The attitude integrated from the
previous star camera reading can be compared to the new reading, and information on the gyroscope rate offset can be deduced by comparing those two measurements, since the integrated
angle is changed when the rate offset changes.  The offset is fitted for every throw and every gyroscope in this manner. This is achieved by modifying the UKF state (called up to now the
attitude because it was holding only the Euler angles) to be a six parameter vector including three Euler angles as well as three offset values, and updating the UKF equations
accordingly. At every time step, the UKF provides an estimate of the six parameter state, as well as an estimate of the 6x6 covariance matrix of the state. The filter is run both
forward and backwards in time. The forward and backward solutions are weighted together to form the average solution, for which the uncertainty is largest mid-throw (between star
camera readings). The weights used to do the average are the inverse of the forward and backward covariances output by the UKF.

\subsection{Least squares optimizer} 

To propagate the star camera attitude using the gyroscope measurements, the transformation matrix between the frame defined by the three gyroscopes in the box and the star camera
frame must be known. An uncertainty on the position of any gyroscope axis will translate into an uncertainty in the reconstructed pointing. By design, the gyroscopes are mounted in
a nearly orthogonal frame. The transformation matrix is broken up into an orthogonalization matrix O, which moves the gyroscopes' true axes from the nearly orthogonal frame to a
truly orthogonal frame, and a rotation matrix R, which rotates this orthogonal frame into the star camera frame. The orthogonalization matrix depends on three misalignment angles,
and the rotation matrix on three rotation angles. Details on conventions used can be found in appendix \ref{appendix: rot and misal}. These parameters also need to be fitted by the
reconstruction software.  Unlike the gyroscope offset, these parameters are assumed to be fixed throughout the flight.  The pointing reconstruction software can attempt to find
those six parameters by minimizing the difference between the integrated attitude and the next star camera reading. This procedure has some degeneracy with finding the gyroscope
bias offset which is broken by rotating around all three gyroscopes during the gyroscope 1/f noise stability timescale. Because those parameters are fixed throughout the
flight, a different method than described for the gyroscope offsets is used. Instead, the UKF is modified to output the differences between the star camera readings and the
integrated attitude during the flight. Note that the difference used is the difference between the star camera reading and the forward integrated solution before the new star
camera reading is incorporated in the solution, in order to provide a non-biased estimate. The differences computed with the backward integrated solutions are also used. This set
of differences is used as a metric to find the correct misalignment and rotation angles. A non-linear least square optimizer (Levenberg-Marquardt) is used on this metric output by
the UKF for the whole dataset. The parameters varied in the optimizer are the six rotation and misalignment angles. The correct step size in the parameter space has to be tuned for
the optimizer to converge.  The UKF is run on data from the entire flight many times through the non-linear least square optimizer. The initial guess to the optimizer, the
parameter space allowed, as well as the results of the reconstruction will be detailed in section \ref{section: results}.

\section{Systematic Errors Affecting Pointing Reconstruction}
\label{section: systematics}

The EBEX scan strategy enables star camera readings only approximatively every 40 seconds.  The designed scan strategy is to raster scan a patch of sky of 20\degree x20\degree,
performing back and forth azimuth throws at a constant elevation before stepping in elevation. The 20\degree azimuth throws are scanned with a constant azimuth velocity of
0.5\degree /s. As discussed previously, star camera images must be taken at velocities smaller than 2'/s, which only happens at scan turnarounds, every 40 seconds.  Between star
camera readings any non-ideality in the star-camera-gyroscope system integrates into a growing pointing error.  Though this pointing system is common on pointed balloon-borne
instruments, the amount of time between star camera images on EBEX requires a refined assessment of the systematic errors such as gyroscope noise, gyroscope gain, or misalignment
between the star camera and each gyroscope. In the following subsections, we list the major systematic errors for our system and we characterize the effect of each error on the
pointing reconstruction error. If relevant, we detail the steps taken in the design of the system or the pointing reconstruction software to minimize the magnitude of the
systematic error. The simulation pipeline detailed in section \ref{section: simulation_pipeline} was written to simulate the relevant systematic effects and ensure that the
reconstructed pointing satisfies the EBEX requirement.

\subsection{Gyroscope Noise} 

\begin{figure}
\centering
\includegraphics[width=3.25in]{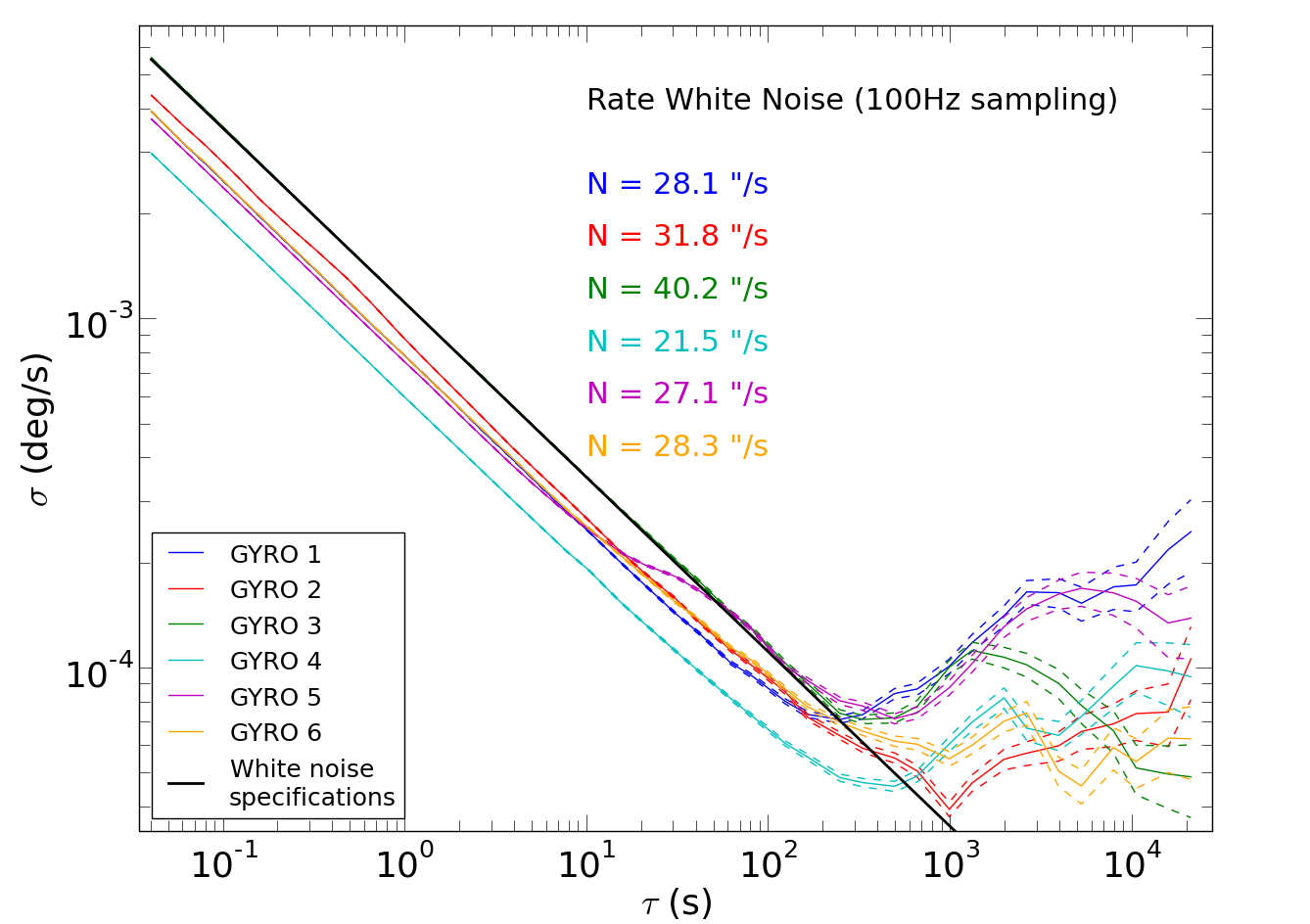}
\caption{Allan variance measurement for all six EBEX gyroscopes. The test is performed by measuring the gyroscope rates for 24 hours while the gondola is immobile on the ground, facing
south (to minimize measurements of the earth rotation rate). For a given chunk length $\tau$, the RMS of the difference in the mean of consecutive chunks is plotted. The white noise
level, measured by the offset of the -1/2 slope on short timescales, is given for each gyroscope and can be compared to the 40"/s specification, plotted in black. The curves' lowest point
indicate the timescales on which the bias varies, roughly 200 seconds.}
\label{fig: allan_variance}
\end{figure}

To assess the effect of gyroscope noise on the reconstructed pointing, we break up the noise into a white noise component and a slow varying drift or bias component. It is common to assess
the noise on different timescales using the Allan variance method \cite{IEEE 952}, which determines the spread in the consecutive means of many chunks of data, as a function of the chunk
length $\tau$. The Allan variance and the white noise specification for each EBEX gyroscope are shown in figure \ref{fig: allan_variance}. 
The uncertainty on the reconstructed attitude coming from integrating gyroscope white noise grows like $\sqrt{N}$, where N is the number of integrated samples. 
The RMS on the reconstructed attitude of a 40 second throw coming from the EBEX gyroscope rate white noise is $\sim$ 11", as is detailed in appendix \ref{appendix: white_noise_rms}. 
This is well under our requirement of RMS = 54", which is why those gyroscopes were chosen.
We then evaluate the effect of bias on the reconstructed attitude.

Figure \ref{fig: allan_variance} shows that the bias is constant on timescales of $\tau \simeq$ 200 seconds. Thus between two star camera readings, the bias will appear as a
constant systematic offset in the gyroscope rate reading.  The gyroscope specifications state that the bias can have values up to 20"/s. Between two star camera readings, a
constant offset of 20"/s in the gyroscope rate will lead to an error in the reconstructed pointing that grows linearly with the number of integrated samples and produces an RMS of
516", which is well above the EBEX requirement.  The pointing reconstruction software therefore needs to fit the bias offset. Details on how this is done were given in section
\ref{section: pointing reconstruction}. 

\subsection{Orientation between the gyroscopes and the star camera}

Six angles are required to rotate three gyroscopes into the star camera frame axis: three angles in the orthogonalization matrix O and three angles in the rotation matrix R. An
uncertainty on one of the angles translates into an uncertainty on the reconstructed pointing. It is not trivial to map how an uncertainty on an angle translates into an error on
the reconstructed pointing, because the mapping is dependant on the actual motion of the telescope. However we can and do simulate it. The error on a reconstructed pointing sample
is proportional to the error in the angle orientation, and to the angular distance traveled while integrating the gyroscopes. A rough estimate is that an error of $\sim$4.5' on the
orientation of a gyroscope translates into an error of 54" in the reconstructed pointing for a 40 second throw. To constrain the misalignment angles, a measurement was made by
rotating the gyroscope box on a rotary table on each of the orthogonal box's side. The ratio between two velocities gives the alignment between two gyroscopes if the box is
perfectly orthogonal. The alignment angles were measured to $\sim$0.3' precision. To ensure the boxes are orthogonal, they are precision machined such that outside surfaces are
mutually parallel and orthogonal with a surface precision of 5 mils, corresponding to a 4.9' angle uncertainty. The misalignment angles are constrained to $\sim$5.2'. This is
larger than the limit allowed by our pointing requirement and so we need to adapt the reconstruction software to fit these angles. Furthermore these measurements constrain the
orthogonalization matrix O, but not the rotation matrix R which changes every time the star camera or the gyroscope box are mounted on the gondola. The rotation matrix R must be
reconstructed using flight data. The procedure to find all six angles with the reconstruction software was detailed in section \ref{section: pointing reconstruction}. The
measurement of the misalignment angles is used in the least squared optimizer to reduce the parameter space of possible angles, ensuring that the optimizer converges. The
simulations detailed in section \ref{section: simulation_pipeline} show that the reconstruction software finds these angles to sufficient accuracy such that the reconstructed
pointing satisfies the EBEX requirement.

Another possible systematic effect is the rotation matrix R slightly changing with elevation. Comparing the measured star camera elevation with the encoder situated on the trunnion
bearing of the inner frame, it was shown that the two measurements spread about an arcminute when the gondola was moved in elevation over its entire 50\degree elevation range. The
gyroscope box A and the star camera 1 being next to each other on the inner frame should however be minimally affected by this effect.

\subsection{Scale factor}

The manufacturer's specifications on the gyroscope scale factors are calculated for the entire +/- 350\degree/s rate range, and are not constraining enough for the regime in which we use
the gyroscopes (+/- 2\degree/s). Therefore the scale factor of each gyroscope is re-measured by aligning it with the gondola's elevation axis and moving the gondola in elevation at various
velocities. The scale factor is found by calculating the best fit line between the encoder and the gyroscope velocities. Scale factors found range between gyroscopes from 1.001 to 1.005 with
a measured uncertainty of 7e-5. The measured value for each gyroscope is used in the pointing reconstruction. The uncertainties on the measurements of the scale factor are used in section
\ref{section: simulation_pipeline} to simulate the effect of errors of this magnitude in order to determine whether or not this characterization is sufficient for the pointing reconstruction
requirements.

\subsubsection{Magnetic fields} 

The fiber optic gyroscopes are sensitive to ambient magnetic fields, which alter the rotation rate reading, so we shielded them by taping them in overlapping strips of Metglas. The
rate dependence on the magnetic field was then tested using a Helmotz coil. The sensitivity to magnetic fields was reduced from 17.2"/s/G to 2.3"/s/G by the Metglas. On the LD
flight trajectory, the earth magnetic field's horizontal component is at the maximum 200 mG. For 20\degree azimuthal throws, this translates to a maximum field change of 68 mG, or
a change in the gyroscope bias of 0.16"/s. As we will show in section \ref{section: results}, this is smaller than the uncertainty to which the bias is fit for every throw and can
be ignored.

\section{The Simulation Pipeline}
\label{section: simulation_pipeline}

A simulation pipeline was developed both to fine-tune the reconstruction software, and to determine the precision with which non-idealities in gyroscope measurements and mount
angles needed to be constrained for the pointing reconstruction to satisfy the EBEX science requirement. The pipeline simulates attitude time streams, called hereafter the parent
pointing.  From these time streams, simulated star camera solutions and gyroscope readings are generated. Systematic effects are also simulated, with orders of magnitude similar to
those measured in section \ref{section: systematics}. As described earlier, the reconstruction software then uses the simulated sensor readings to reconstruct attitude time
streams. In this section we detail how sensor data and systematic errors are simulated.  Results on the pointing reconstruction for simulated data are available in section
\ref{section: results}.

\subsection{Simulated Parent Pointing} 

Simulating the parent pointing is the first step to generating simulated sensor data. The parent pointing consists of a time stream where each sample provides information to orient
the telescope in the equatorial reference frame: RA, Dec, Roll.  Before the LD flight, the parent pointing was created by simulating raster scanning with a given scan speed,
azimuth throw amplitude, and elevation step size. After the LD flight, a more realistic parent pointing was generated. We focus on post LD flight simulations in this paper.  The
parent pointing used is a smoothed, rough estimate of the 2012/2013 EBEX flight pointing, first using real-time in-flight pointing and after a first iteration using the
reconstructed pointing. The star camera and gyroscopes are simulated for the entire 11 day flight. The sensor data generated has the same format as real data from flight. This
simulated sensor data is then used as input to the pointing reconstruction software. The attitude time streams output by the reconstruction software can then be compared to the
input parent pointing to estimate the quality of the reconstruction. This is done in section \ref{section: results}.

\subsection{Simulated Sensor Data} 

\subsubsection{Star Camera Solutions} 

A star camera solution consists of three Euler angles (RA, Dec and Roll) at a given time step corresponding to when the image was taken (the trigger).  The first step to simulate
star camera solutions is to simulate the triggers. The star cameras only take pictures when conditions of slow velocity and acceleration are met. The same code used in FCP to
determine when the star cameras should take images can be used to place the triggers in the simulated data. Alternatively, the code can use actual triggers from flight. The results
in this paper are based on the latter method. For a given trigger, the star camera solution is generated by adding Gaussian noise to the parent RA, Dec and Roll at that time step.
The magnitude of the Gaussian noise is given by the accuracy of the star camera solutions.  Before flight, the magnitude of the star camera errors was determined by simulating
stars and sky brightness, and solving the simulated images. After flight, the magnitude of the star camera images was determined using the measured errors from flight, as shown in
figure \ref{fig: star camera solutions}.

\subsubsection{Gyroscope rates} 

The gyroscope rates $\omega_1, \omega_2, \omega_3$ need to be computed from the parent time streams. The process for computing simulated gyroscope rates is equivalent to inverting
equation \ref{eq: propagation}: for every time step, the parent Euler angles, parent Euler angle rates, and the R and O matrices are used to compute the gyroscope rates. Several
systematic errors are then included when simulating the rates: additive noise, gains, misalignment and rotation angles. The aim is to simulate the systematic errors with a
realistic magnitude, and verify that the reconstruction software is capable of reconstructing the pointing well enough to satisfy the EBEX requirement. 

We now detail how each systematic error is simulated.  The misalignment and rotation angles are the parameters on which the matrices O and R depend. The measurement of the
misalignment angles described in section \ref{section: systematics} showed that all misalignments are smaller than 0.5\degree. For each simulation, the misalignment angles are thus
chosen from a Gaussian distribution with mean 0\degree and standard deviation 0.5\degree.  The rotation angles are similarly chosen from a Gaussian distribution with mean 0\degree
and standard deviation 10\degree, which is the measured uncertainty on the rotation between the gyroscope boxes and the star camera pressure vessels coming from their respective
mounting hardware. The rates $\omega_1, \omega_2, \omega_3$ are then multiplied by a scale factor to simulate unknown gains. The scale factor is randomly chosen from a Gaussian
distribution with a mean of 1.0 and a standard deviation of 7e-5, corresponding to the uncertainties on the measurements of the scale factors described in section \ref{section:
systematics}. Note that unlike other effects simulated, the scale factors are not found by the reconstruction software.  Instead we used simulations to show that the uncertainties
on the measured gains were small enough that the reconstructed pointing satisfies the EBEX requirement without further processing necessary.  Finally we add the gyroscope noise. To
simulate realistic noise, the power spectra of the EBEX gyroscopes are computed using the same data taken for the Allan variance measurements. Each power spectrum P is fitted to
the model:

\begin{equation}
P = n (1 +  (f_{knee}/f)^{\alpha})
\end{equation}

where n is the white noise level and $f_{knee}$ the critical frequency for drifts, seen to be roughly 5e-3 Hz from figure \ref{fig: allan_variance}. The $\alpha$ values for EBEX gyroscopes
vary between 1 and 1.5. The gyroscope noise is then created by generating a time stream of random Gaussian noise with standard deviation n, taking the power spectrum of this white noise,
scaling the lower frequencies according to $(f_{knee}/f)^{\alpha}$, and taking the inverse power spectrum. The noise time stream is then added to the gyroscope rates.

\section{The Long Duration Flight}
\label{section: flight}

\begin{figure}
\centering
\includegraphics[width=3.5in]{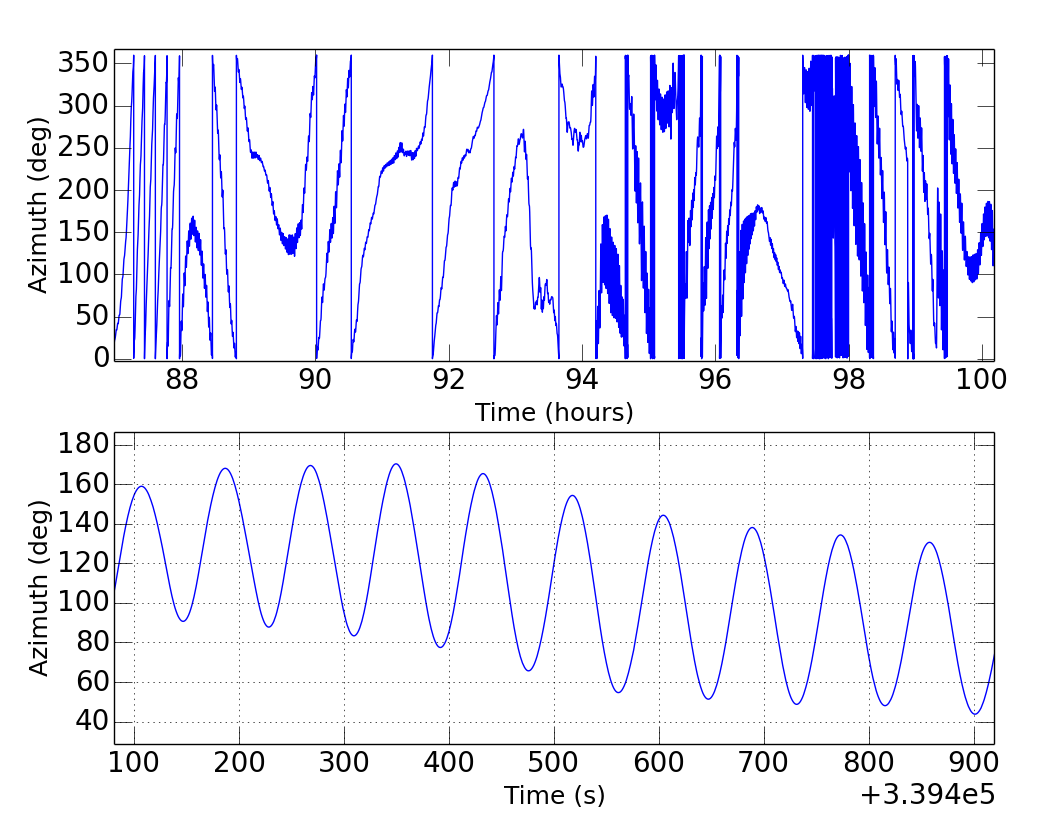}
\caption{Typical patterns in the azimuth motion during the LD flight. The top panel shows the long term drifts. The bottom panel shows the 80 second oscillation on top of the long
term drifts. Star camera readings are taken every $\sim$40 seconds when the gondola changes azimuth direction.}
\label{fig: az motion}
\end{figure}

\begin{figure}
\centering 
\includegraphics[width=3.3in]{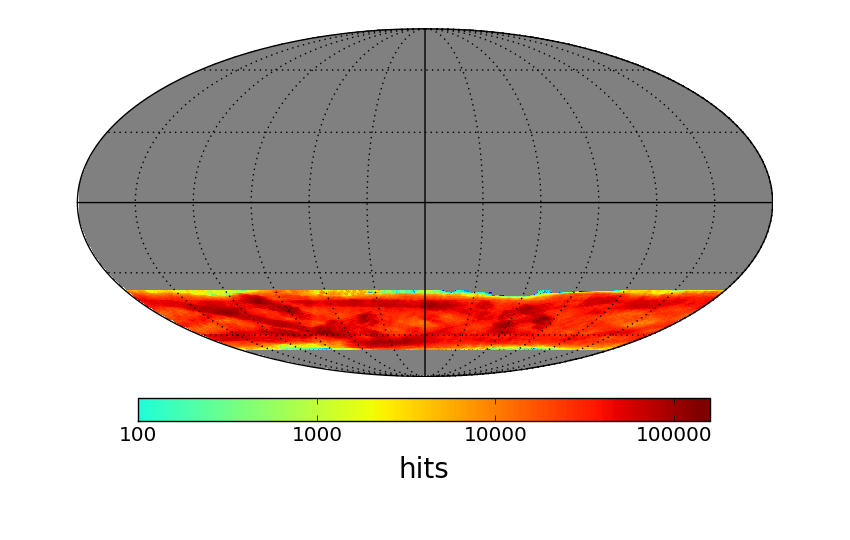} 
\caption[] {A map in equatorial coordinates of the number of detector samples per pixel for the 2012/2013 EBEX flight from
all frequency bands. HEALPix \cite{healpix}  was used for the pixelization scheme and the pixels are approximately 6.9' on a side. This pixel size is 16 times larger
than the size used to determine the pointing requirements. A single full rotation of the telescope boresight in azimuth gives a sinus-wave line bounded by the edges of the Dec strip. Full
coverage of the strip is achieved because the phase of this sinusoid changes throughout the day as the Earth rotates and because the focal plane is about 6\degree across in both azimuth and 
elevation.}
\label{fig: hitmap}
\end{figure}

EBEX launched from McMurdo, Antarctica on December 29th 2012. It circumnavigated the continent taking data during the first 11 days, at an altitude of $\sim$ 35 km.
Shortly after launch we discovered that the azimuth motor controller was overheating and causing a motor controller shutdown. The problem was traced to an error in the thermal design of the motor
controller mounting. Without active control, the azimuth of the gondola was determined by the rotation of the balloon and the rotational spring constant of the flight train. The resulting
motion was composed of slow full rotation (approximately 20 min/rotation) and a faster $\sim$80 second period sinusoid motion in azimuth of 50\degree - 100\degree peak-to-peak. Figure
\ref{fig: az motion} shows typical short and long timescale behavior of the azimuth in flight.  With this type of azimuth motion we decided to maintain the gondola at constant elevation of
54\degree, chosen to maintain angular distance of $\sim$15\degree from the balloon above, and from the sun at its maximum elevation below. A map of the pointing of all detector samples (for
all frequency bands) is shown in figure \ref{fig: hitmap}. EBEX scanned a strip of approximately constant Dec between -67.9\degree and -38.9\degree. Counting all 6.9' pixels that
have more than 10\% of the maximum number of hits/pixel we find a total scanned area of 5735 square degrees. By a fortunate coincidence, the natural oscillating frequency of the gondola,
80 seconds, matched exactly the frequency of the design scan strategy: the gondola came to a stop every $\sim$ 40 seconds, enabling the star cameras to take pictures. In this manner, all the
pre-flight work assessing the gyroscope noise and non-idealities was still relevant to the actual scan pattern of the LD flight. 

\subsection{Performance of the primary sensors}

\begin{figure}
    \centering
    \includegraphics[width=3.25in]{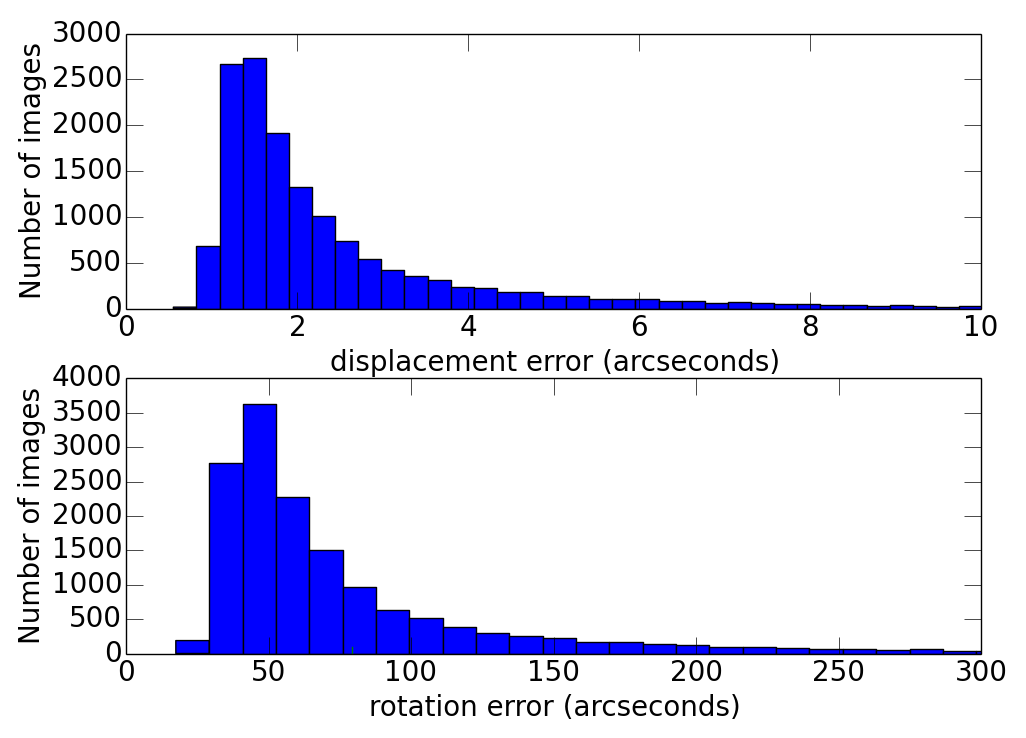}
    \caption{Histogram of the pointing solution uncertainty for all solved images from star camera 0.}
    \label{fig: star camera solutions}
\end{figure}

The star cameras performed well during flight, consistently solving the images real-time with minimal intervention. The STARS software overcame several unanticipated challenges,
detailed in this paragraph. The loss of azimuth control prevented us from performing the autofocus algorithm which requires stationary pointing, and as a result both star cameras
were slightly out of focus during the entire flight. However the STARS software continued to consistently find the stars in the images because of its robust source detection
algorithm. To solve images, the star cameras normally use a pointing guess from coarse sensors, in order to minimize the search radius when matching the stars in the image to the
catalog of stars. The coarse sensors provide a guess in the horizontal reference frame, and this guess is translated in the equatorial reference frame using the GPS latitude,
longitude and the time. The GPS failed for multiple sections of the flight, which prevented the pointing guess to be rotated in the equatorial frame in which the star cameras
operate. In those sections the star cameras continued finding solutions in several seconds, switching to lost-in-space mode. This capacity to solve the images in several seconds
without any guess was made possible in STARS by optimizing the catalog of stars for the EBEX star cameras \cite{chappy spie}. The STARS software continued finding stars through
different image non-idealities encountered in flight: increased atmospheric brightness from the high elevation, polar mesospheric clouds, vignetting and reflections from the CCD. A
detailed review of the performance of the star cameras can be found in the Chapman et al. companion paper \cite{chappy ieee paper}. The two star cameras combined took 41,262
images, $\sim$80\% of which solved. Most of the remaining images are either saturated or within 30\degree of the sun. Figure \ref{fig: star camera solutions} shows a histogram of
the solution errors for all star camera 0 flight images as reported by the pattern matching least squared algorithm. The gyroscopes also performed well, recording data continuously
and exhibiting behavior in accordance with pre-flight measurements.

\section{Pointing Reconstruction Results} 
\label{section: results}

This section provides details on running the pointing reconstruction software, describes the methods available to evaluate the reconstructed pointing and gives the results obtained both from
simulations and from the LD flight data. 

\subsection{Running the pointing reconstruction software} 

The reconstruction software is composed of the UKF and the least squared optimizer. The UKF takes as input star camera solutions and associated uncertainties, gyroscope rates and estimated
white noise, bias timescales, and misalignment and rotation angles. It outputs time streams of Euler angles (RA, Dec and Roll) and gyroscope bias with their uncertainties. It also outputs the
metric used by the optimizer.  To run the reconstruction software, the least square optimizer needs an initial guess for the parameters (misalignment and rotation angles), as well as the
size of the parameter space (how big the angles can get). The initial guess for all rotation and misalignment angles is zero, and the parameter space allowed is 0.5\degree for misalignment
angles, and 10\degree for rotation angles. The correct step size in the parameter space has to be tuned for the optimizer to converge and find the correct angles.  This is done by
trial and error using simulated data. After tuning the step
size, it takes on average 90 iterations for the optimizer to converge, and each iteration (reconstructing the pointing on the entire 11 day flight), takes approximatively 80 minutes on a
single 2.1 GHz processor. The actual code was multi-processed. The UKF is then re-run one last time with the optimized rotation and misalignment angles. The output pointing time streams and
pointing uncertainties are written to disk. The simulations show us that the rotation and misalignment angles are found to within $\sim$ 0.001 radians on average, and are more accurately found if
data with varied motion are input to the UKF. It is possible to run the optimizer with larger values allowed for the rotation angles (up to tens of degrees). With the proper step size, the
optimizer will still converge, though it takes more iterations.

\subsection{Evaluating the reconstructed pointing} 

There are three methods available to evaluate the quality of the reconstructed pointing: the UKF estimation method, the star camera difference method, and the parent difference
method . The third method presented is only available for simulated data. The goal is to estimate the error of the reconstructed pointing over a 40 second throw. The error over the
throw is the square root of the average of all variances across all samples between two star camera readings separated by 40 seconds (RMS). The estimated error on the throw can
then be compared both to the pure white noise case (RMS = 11") and to the EBEX requirement (RMS = 54").

\subsubsection{Estimated error output by the UKF} 
At every time step, the UKF outputs a covariance matrix along with an estimate of the state. This means that an estimated
uncertainty for RA, Dec and Roll is available at every time step.  The UKF estimates the covariance using several sources of information: the gyroscope white noise value (input by
the user), the bias stability timescale (input by the user), the star camera uncertainty (input by the user), and the measurements of the difference between the integrated pointing
with the new star camera reading. To summarize, it propagates the uncertainty given the input noise, and updates the uncertainty when new measurements are available. The RMS on a 40
second throw can be calculated by averaging all these UKF estimated variances for every sample in a 40 second throw:

\begin{equation}
\label{eq: rms}
RMS = \sqrt{\frac{1}{T}\sum_{t=0}^{t=T}\sigma_t^2dt}
\end{equation}

where T=40 seconds is the throw size in time and $\sigma_t$ is the uncertainty of the reconstructed pointing output by the UKF for sample t. Included in the average are all uncertainties on
the Dec angle, and all uncertainties on the RA angle corrected by cos(Dec) to represent the angular distance on the sky. Hereafter we call RA*cos(Dec) the cross-Dec. To note is that the UKF
estimate is heavily dependant on the values input by the user regarding gyroscope white noise, bias timescales and star camera noise.

\subsubsection{Comparing the forward and backward solutions to the star camera readings} 

At every star camera reading, we can compare the pointing integrated from the previous star camera reading to the new star camera solution. This can be used to measure the quality
of the reconstruction. To provide a non biased estimate of the error, the difference should be calculated before the new reading is incorporated in the integrated pointing.  This
can be done with the forward integrated pointing and with the backward integrated pointing (but not with the averaged pointing). We need to link these star camera differences to
the RMS over a 40 second throw, which is the average variance of all samples in the throw. Though the majority of star camera readings are separated by 40 seconds, there are many
star camera readings separated by any time difference in the range 0 to 40 seconds. Each time there is a star camera at time t away from the previous star camera reading, the
difference between the forward and backward integrated solution with the star camera reading gives us and indication of the typical error at this time t. We can bin all throws in
2.5 second bins, and collect for the entire flight all the star camera differences that fall within each bin. For each bin corresponding to time t, the variance of all the
differences in that bin gives us an estimate of the variance $\sigma_{t_F}^2$ or $\sigma_{t_B}^2$ at time t. Here F and B refer to the forward and backward variances. The average
variance at time t can be obtained easily from them:

\begin{align}
\label{eq: xsc rms}
\frac{1}{\sigma_t^2} &= \frac{1}{\sigma_{t_F}^2} + \frac{1}{\sigma_{t_B}^2} \\
                     &= \frac{1}{\sigma_{t_F}^2} + \frac{1}{\sigma_{T-t_F}^2} \\
\sigma_t^2 &= \frac{\sigma_{t_F}^2 \sigma_{T-t_F}^2}{\sigma_{t_F}^2 + \sigma_{T-t_F}^2}
\end{align}

where T=40 second is the period of the throw. Then the RMS on the 40 second throw is simply the square root of the average of all $\sigma_t^2$, as laid out in equation \ref{eq:
rms}. This method to estimate the RMS will be called the "star camera difference method" in the following subsections. In is interesting to note that the $\sigma_{t_F}$ and
$\sigma_{t_B}$ will grow like $\sqrt{t}$ if the data is dominated by gyroscope rate white noise. If the data is dominated by a bias offset systematic error, the uncertainties will
grow like $t$. Finally if the data is dominated by a gyroscope angle systematic error, the error will grow like the distance traveled since the last star camera reading. This
method is illustrated by plotting the $\sigma_{t_F}$ in figures \ref{fig: simul error grow} (for simulated data) and \ref{fig: flight_reconstruction} (for LD flight data).

\subsubsection{Comparing the parent pointing to the reconstructed pointing (simulations only)}

This method is only available for simulated data and is the best possible estimate for the RMS. For simulated data, the parent pointing is available for every sample. For every
sample in a 40 second throw, the difference between the averaged reconstructed pointing and the parent pointing is calculated. The RMS of all these differences is by definition the
RMS on the throw. The histogram of these differences is plotted in blue in figure \ref{fig: simul_histograms}. This measurement is also used to validate the star camera difference
method by showing that the two estimations give similar result for simulated data.

\subsection{Results for simulations}

\begin{figure*}
\centering
\includegraphics[width=7.5in]{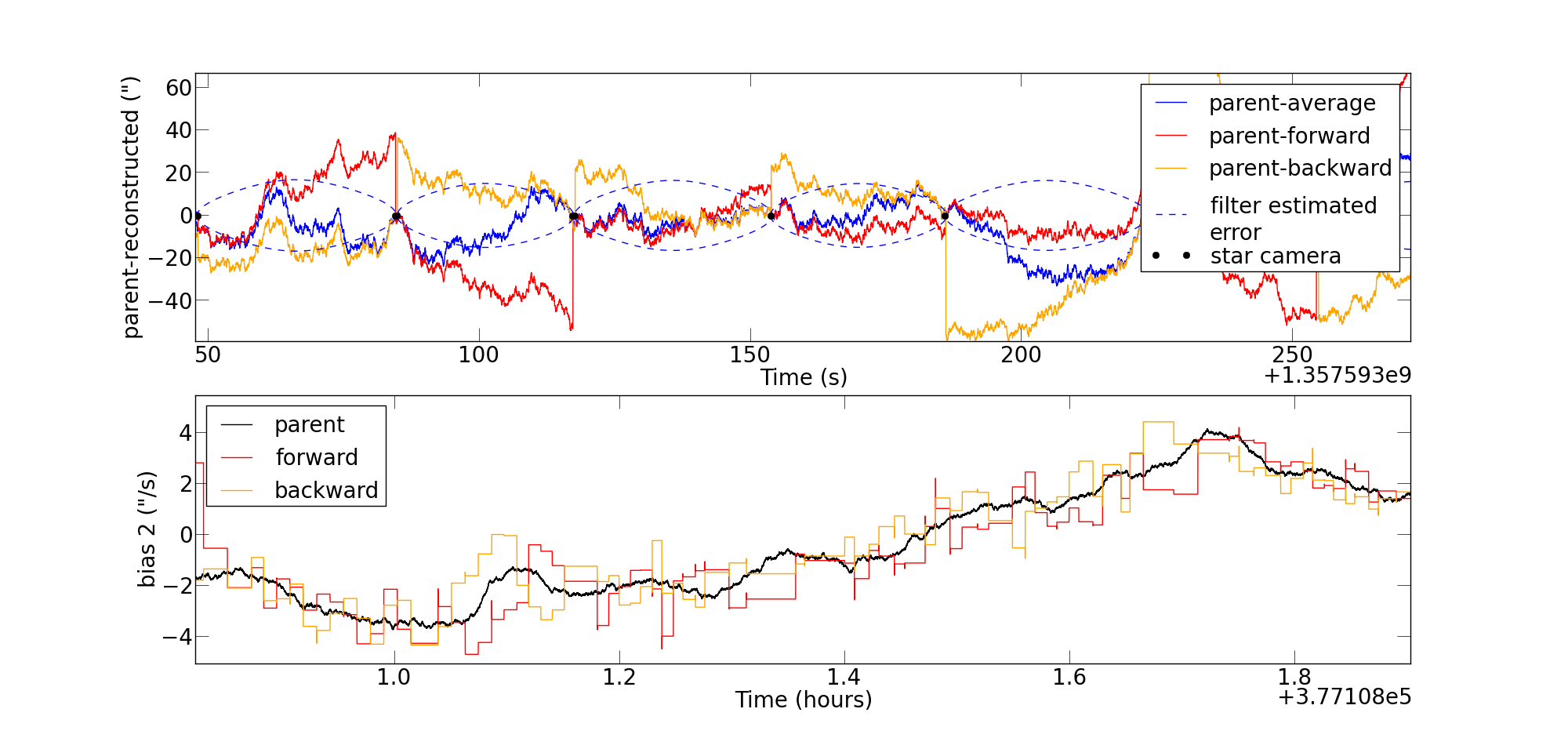}
\caption{{\bf{Top panel}}: Difference between parent and reconstructed time streams of Dec angle (forward: red, backward: orange, average: blue). Roughly 5 throws are shown.
The black dots corresponds to the star camera Dec solutions. In dashed blue is the UKF estimated uncertainty on the average solution.
{\bf{Bottom panel}}: The parent (black) and reconstructed 
(red and orange) gyroscope bias are plotted for gyroscope 2. Note that this plot shows much longer timescales than the top plot. The parent bias plotted corresponds to the parent gyroscope noise, 
averaged with a 200 seconds window size. The reconstructed bias is calculated at every star camera reading, hence the step pattern on the plot. The forward and backward biases are
reconstructed to within a few arcseconds. The average bias is not plotted as it is not actually used in the reconstruction: the forward pointing uses the forward bias, and the backward pointing
the backward bias. The average solution is then computed by averaging the forward and backward pointings.}
\label{fig: simul_timestreams}
\end{figure*}

Figure \ref{fig: simul_timestreams} shows time streams of the difference between parent and reconstructed Dec angle, as well as parent and reconstructed gyroscope bias for one of
the gyroscopes. The average pointing time stream is the weighted average of the forward and backward reconstructed time streams. The weights used are the inverse of the variances
output by the reconstruction software for every sample. The UKF estimated error for the average solution is also plotted in dashed blue for every sample, and it is apparent that
on average the error is largest mid-throw in between two star camera readings. The bias is reconstructed to within $\sim$1", which is sufficient to satisfy the
EBEX requirement.

\begin{figure}
\centering
\includegraphics[width=3.5in]{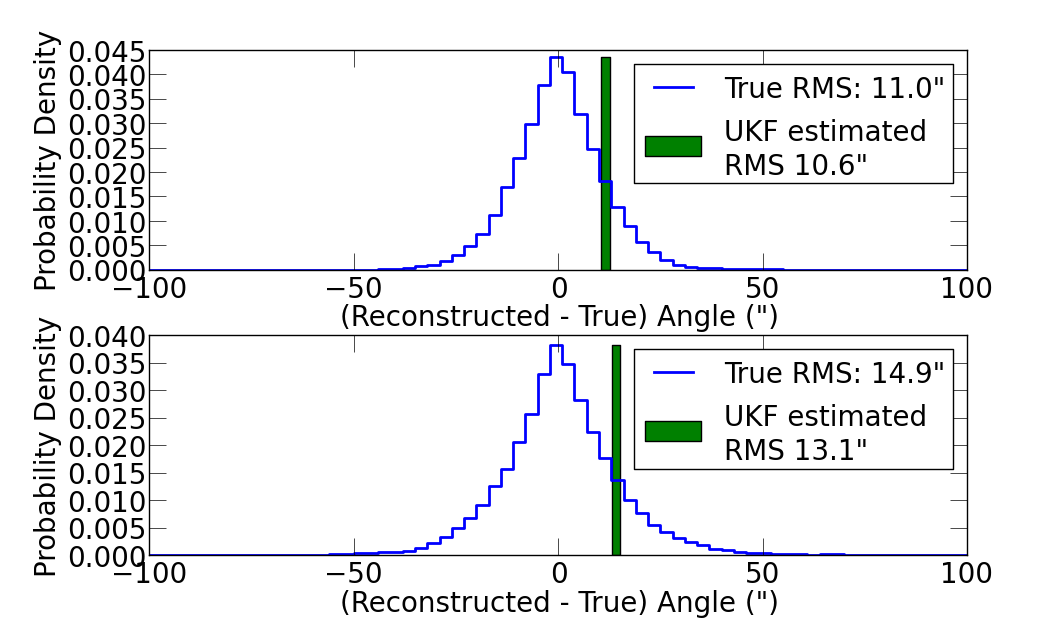}

\caption{Histograms of the difference between parent and reconstructed pointing, for all samples in all 40 second simulated throws. {\bf{Top}}: White noise reference case. The
simulated data includes only gyroscope and star camera white noise. The reconstruction software doesn't fit the gyroscope bias or the gyroscope angles. The measured RMS, 11",
matches the theoretical case calculated in appendix \ref{appendix: white_noise_rms}. The RMS estimated by the UKF is shown in green.  {\bf{Bottom}}: Full noise case. The simulated
data includes gyroscope bias, misalignment and rotation angles, and non-unity gains. The reconstruction software fits the gyroscope bias, misalignment and rotation angles. The RMS
is smaller than the EBEX requirement of 54", though the RMS is larger in the full noise simulation than in the white noise case, as expected.}

\label{fig: simul_histograms}
\end{figure}

To characterize the quality of the reconstruction on the simulated data, the "parent difference" method is used. Histograms of the difference between the simulated parent and the
reconstructed Dec and cross-Dec are plotted in blue in figure \ref{fig: simul_histograms}, for throws that are 40 +/-1 second long. Both the pure white noise case and the full
noise case are plotted.  The simulation with pure gyroscope white noise has a measured RMS of 11", which is in accordance with the theoretical prediction (calculation detailed in
appendix \ref{appendix: white_noise_rms}) confirming that the reconstruction software behaves as expected.  Note that even for the  pure gyroscope white noise case, these
histograms are not expected to be Gaussian, as the attitude along the throw is sampled from Gaussian distributions with growing standard deviation, and therefore the final
histogram is the sum of Gaussian distributions with different standard deviations.  For the full noise case, the RMS of the entire flight simulation for all 40 second throws,
14.9", is smaller than the required RMS of 54", ensuring our pointing reconstruction requirements are met in a simulation with realistic noise and systematic errors. This
measurement validates that there is no need to measure the gains to better accuracy or to account for the leftover susceptibility to electromagnetic fields. It also confirms that
the bias and gyroscope orientation are found to sufficient accuracy by the reconstruction software. The RMS estimated by the UKF is shown in green.

\begin{figure}
\centering
\includegraphics[width=3.3in]{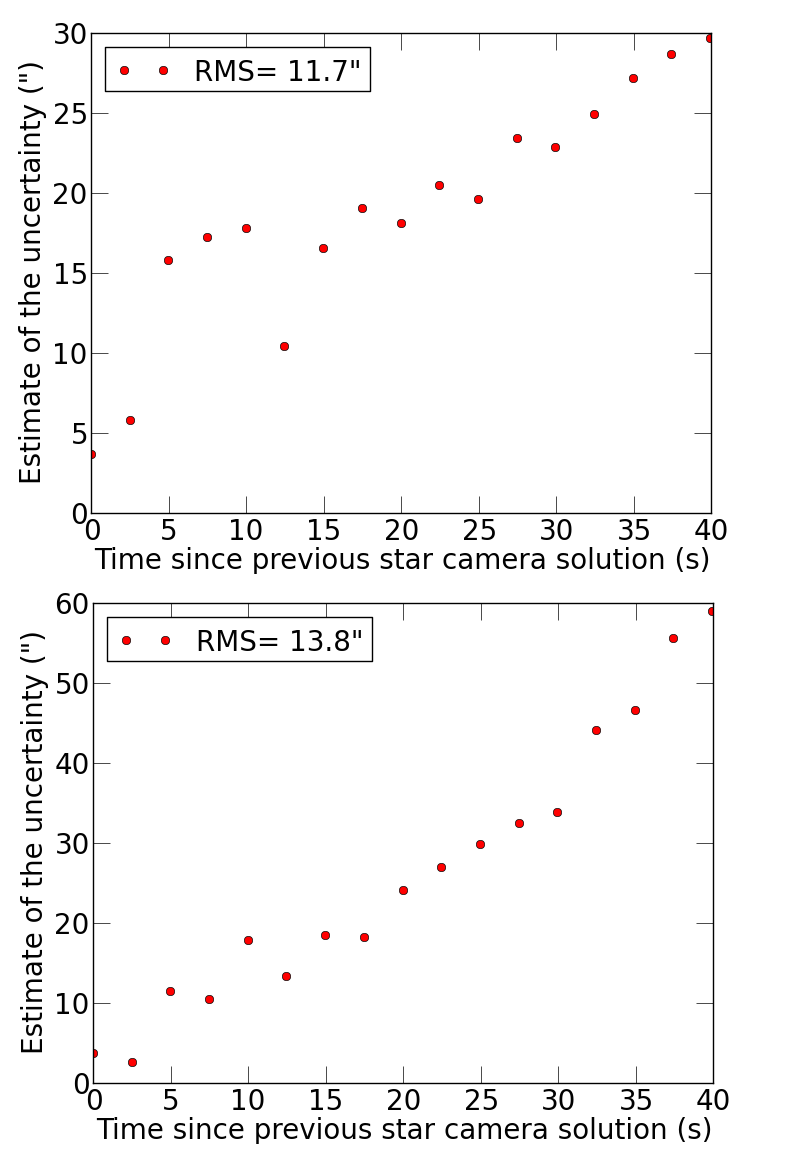}

\caption{Star camera difference method applied to simulations. The red dots are the estimate of $\sigma_{t_F}$ and $\sigma_{t_B}$, the forward and backward errors as a function of
the distance in time to the previous star camera reading. All throws are binned in 2.5 second bins. Each dot is computed by collecting the differences between star camera solutions
and forward and backward pointing for all star camera images that fall within than bin. The dot is then the standard deviation of all differences in that bin.  The RMS on the throw
is estimated from those points using equations \ref{eq: xsc rms} and \ref{eq: rms}. This RMS, which is computed here for simulated data without any reference to the parent
pointing, can be compared to the true RMS computed with the "parent difference" method in figure \ref{fig: simul_histograms}. {\bf{Top:}}White noise only case.  {\bf{Bottom:}} Full
noise simulation.}

\label{fig: simul error grow}
\end{figure}

To validate the "star camera difference" method, the method is applied on simulated star camera solutions and reconstructed pointing from simulations. Note that this method does not
require knowledge of the parent pointing.  Figure \ref{fig: simul error grow} shows the estimated $\sigma_{t_F}$ error as a function of the distance $t$ to the star camera
reading. For each simulated star camera image, the difference is computed between the star camera Dec and Cross-Dec, and the forward and backward reconstructed Dec and cross-Dec.
The differences are binned as a function of the distance to the previous star camera reading, in bins of 2.5 seconds. For each bin, the standard deviation of the
differences is plotted. The RMS on the throw is obtained from those points using equations \ref{eq: xsc rms} and \ref{eq: rms}. This is done both for the white noise simulation and
for the full noise simulation cases.  The RMS found in each case, 11.7" and 13.8", are to be compared to the true RMS obtained from the "parent difference" method illustrated in
    figure \ref{fig: simul_histograms}: 11" and 14.9".  This confirms the validity of the "star camera difference method" to estimate the true RMS over the throw.

\subsection{Results for LD flight data}

\begin{figure}
\centering
\includegraphics[width=3.3in]{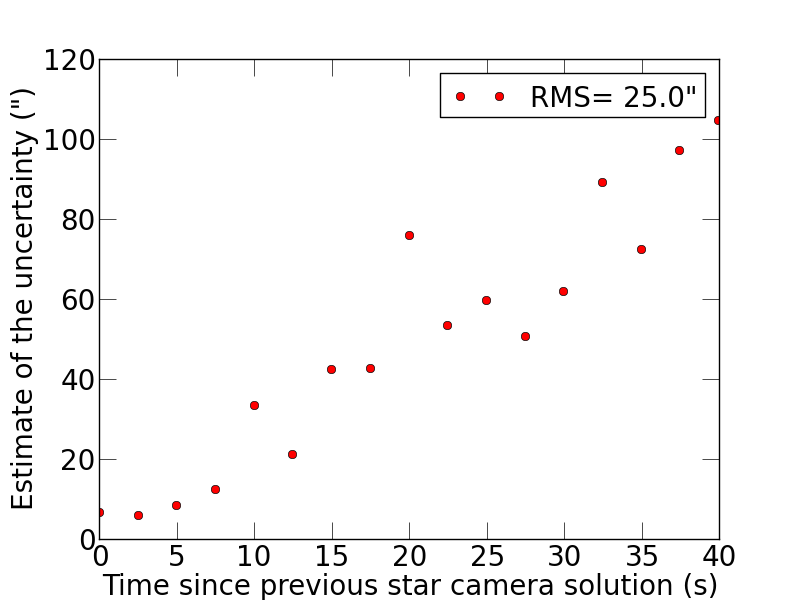}

\caption{Star camera difference method applied to the 2012/2013 EBEX flight data. The red dots are the estimate of $\sigma_{t_F}$ and $\sigma_{t_B}$, the forward and backward errors as a function of
the distance in time to the previous star camera reading. All throws are binned in 2.5 second bins. Each dot is computed by collecting the differences between star camera solutions
and forward and backward pointing for all star camera images that fall within than bin. The dot is then the standard deviation of all differences in that bin.  The RMS on the throw
is estimated from those points using equations \ref{eq: xsc rms} and \ref{eq: rms}.}

\label{fig: flight_reconstruction}
\end{figure}

For simplicity only results with gyroscope Box A and star camera 0 are presented here. The gains measured pre-flight are used in the reconstruction software.  The "star camera
difference" method is applied to the 2012/2013 LD flight reconstructed pointing in figure \ref{fig: flight_reconstruction}. In the figure, the forward $\sigma_{t_F}$ and backward
$\sigma_{t_B}$ estimated error are plotted as a function of the distance $t$ to the star camera reading. The plot is constructed identically to figure \ref{fig: simul error
grow} in the previous subsection, but using LD flight data instead of simulated data.  Using the points in figure \ref{fig: flight_reconstruction}, we use equations  \ref{eq: xsc
rms} and \ref{eq: rms} to estimate the RMS on a 40 second throw. The result is RMS = 25", a factor of two under the EBEX pointing reconstruction requirement of 54". It is
interesting to note that the UKF reports an RMS of only 11.7". The fact that the estimated RMS for real data is higher than for simulations and that the UKF finds an underestimate
of the error suggests an un-modeled and previously unmeasured source of gyroscope noise (the UKF estimate is highly dependant on the user input noise magnitude, and the UKF
estimated error is smaller for smaller input noise values). Finally, we transform this measured RMS back to an error in the map domain and estimate this error to be 4.6" (see
appendix \ref{appendix: transform requirements}).

\section{Conclusion}

A pointing system for fast-scanning balloon platform was presented, which enables in-flight and post-flight pointing reconstruction. For a gyroscope and star camera system, periods of tens
of seconds between star camera images require careful control of the systematic errors. For an average throw length of 40 seconds, an RMS of 25" is achieved by designing a system
minimizing possible non idealities,  measuring all gyroscopes non-idealities pre-flight, and using a reconstruction software able to find gyroscope bias rates, as well as rotation and
misalignment angles. The development of a realistic simulation pipeline was crucial in the development, understanding and fine-tuning of the reconstruction software. The growing availability
of pre-orthogonalized, low noise gyroscope systems extends the methods described in the paper to longer throw lengths.

\acknowledgments EBEX is a NASA supported mission through grant numbers NNX08AG40G and NNX07AP36H. We thank Columbia Scientific Balloon Facility for their enthusiastic support of EBEX, and
the BLAST team for sharing their ACS system and knowledge. We also acknowledge support from NSF, CNRS, Minnesota Super Computing Institute, Minnesota and Rhode Island Space Grant Consortia,
the Science and Technology Facilities Council in the UK, Sigma Xi, private donations, and funding from collaborating institutions.  This research used resources of the National Energy Research Scientific
Computing Center, which is supported by the office of Science of the U.S. Department of Energy under contract No. DE-AC02-05CH11231. The McGill authors acknowledge funding from the Canadian
Space Agency, Natural Sciences and Engineering Research Council, Canadian Institute for Advanced Research, Canadian Foundation for Innovation and Canada Research Chairs program. Some of the
results in this paper have been derived using the HEALPix package.

\bibliographystyle{IEEEtran}

\appendices               

\section{Estimating the error on a single throw from gyroscope noise}  
\label{appendix: white_noise_rms}

To assess the quality of the pointing reconstruction, we use the RMS error over a throw. It is interesting to estimate what is the expected RMS on a throw given different
scenarios: first, gyroscope rate white noise only, then adding systematic errors.  Let us first consider the one dimension case, where the gyroscope noise is white and the
gyroscope is perfectly aligned with the direction of motion.  The gyroscope measures a rate $\omega$, and the equation of motion is:

\begin{equation}
\label{eq: simple motion}
\dot{\theta} = \omega \Delta t
\end{equation}

The variance on the integrated angle from a single time step is :

\begin{align}
\sigma^2_{\theta_{k+1}} &= \sigma^2_{\theta_k} + \sigma_{\omega}^2 \Delta t^2 \\
                       &= \sigma^2_{\theta_k} + \sigma_{\theta_0}^2
\end{align}

The gyroscopes have a rate white noise of $\sigma_{\omega} = 40.0"/s$, and the time step is $\sim$ 0.01 seconds, so the uncertainty on the angle from a single step of integration
is $\sigma_{\theta_0} = 0.4"$.  N time steps after a star camera, the integrated pointing in a given time direction has a variance:

\begin{equation}
\label{eq: one direction uncertainty}
\sigma_{\theta_N}^2 = N\sigma_{\theta_0}^2 + \sigma_{sc}^2 
\end{equation}

where $\sigma_{sc}$ is the uncertainty on the star camera solution. At any point in the scan, the reconstruction is the
weighted average of the forward and backward solutions. If M is the total number of samples between two star camera readings, the variance of the pointing at time step N is:

\begin{align}
\label{eq: forward_backward}
\frac{1}{\sigma_{\theta_N}^2} &= \frac{1}{\sigma_{\theta_N}^{2^{forward}}} + \frac{1}{\sigma_{\theta_{M-N}}^{2^{backrward}}} \\
                              &= \frac{1}{N \sigma_{\theta_0}^2 + \sigma_{sc}^2} + \frac{1}{(M-N) \sigma_{\theta_0}^2 + \sigma_{sc}^2} \\
                              &= \frac{M\sigma_{\theta_0}^2 + 2\sigma_{sc}^2 }{N (M-N) \sigma_{\theta_0}^4 + M \sigma_{\theta_0}^2 \sigma_{sc}^2 + \sigma_{sc}^4}
\end{align}

The RMS on the throw can be expressed as the average of all the variances across the throw:

\begin{align}
\label{eq: rms integration}
RMS &= \sqrt{\frac{1}{M} \sum_{N=0}^{N=M} \sigma_{\theta_N}^2} \\
             &= \sqrt{\frac{1}{M} \sum_{N=0}^{N=M} \frac{N (M-N) \sigma_{\theta_0}^4 + M \sigma_{\theta_0}^2 \sigma_{sc}^2 + \sigma_{sc}^4}{M\sigma_{\theta_0}^2 + 2\sigma_{sc}^2 } } \\
             &= \sqrt{\frac{\frac{1}{6}M^2 \sigma_{\theta_0}^4 + M \sigma_{\theta_0}^2 \sigma_{sc}^2 + \sigma_{sc}^4}{M\sigma_{\theta_0}^2 + 2\sigma_{sc}^2}} \\
             &\sim \sigma_{\theta_0} \sqrt{\frac{M}{6}}
\end{align}

where we neglected the star camera errors $\sigma_{sc}$ as they are small compared to gyroscope error.
On a 40 second throw, M$\sim$4000, and thus RMS$\sim$11".

For the three dimension case, the equation of motion (eq. \ref{eq: simple motion}) needs to be replaced with equations (\ref{eq: propagation}), and the uncertainty of each
integrated Euler angle after N steps now depends on the specific motion and values of all the angles across the throw. However with our typical velocities and uncertainties, the
1-D calculation is a good approximation.

Systematic errors, primarily incorrect bias offsets or incorrect misalignment and rotation angles found by the UKF can also contribute to the error on the reconstructed pointing.  
This class of systematic errors grows linearly with time.  Let us call this error $\sigma_S$. The error on the attitude N samples after a star camera reading is now:

\begin{equation}
\label{eq: real_sigma_sample_N_one_direction}
\sigma_{\theta_N} = \sqrt{N} \sigma_{\theta_0} + N \sigma_S
\end{equation}

Integrating forward and backward similarly to equation \ref{eq: forward_backward} , the weighted average at sample N has uncertainty:

\begin{align}
\label{eq: real_sigma_sample_N}
\frac{1}{\sigma_{\theta_N}^2} &= \frac{1}{\sigma_{\theta_N}^{2^{forward}}} + \frac{1}{\sigma_{\theta_{M-N}}^{2^{backrward}}} \\
                              &= \frac{1}{(\sqrt{N} \sigma_{\theta_0} + N \sigma_S)^2} \\
                              &+\frac{1}{(\sqrt{M-N} \sigma_{\theta_0} + (M-N) \sigma_S)^2} 
\end{align}

\section{Transforming requirements from map domain to error on a single scan}        
\label{appendix: transform requirements}

The requirements for the pointing reconstruction on EBEX are set in the map domain. The average of all pointing errors within a pixel must be smaller than $\sim$10". We want to map this
average error in a pixel to an error on the RMS of a throw. If a pixel is visited P times from independent scans, the sum of all the pointing errors has a variance $\sigma_{sum}^2$ :

\begin{equation}
\sigma_{sum}^2 = \sum_{i=1}^{i=P} \sigma_{\theta_i}^2
\end{equation}

where we made the reasonable assumption that the errors in between scans are uncorrelated. Each time the pixel is hit, it is from a random position in the throw, and so we can compare the
previous equation to the RMS equation:

\begin{equation}
RMS^2 =  \frac{1}{M}\sum_{t=0}^{t=M}\sigma_{\theta_t}^2
\end{equation}

to deduce that $\sigma_{sum}$ is in fact related to the RMS by $\sigma_{sum} = \sqrt{P} \cdot RMS$. Finally we are interested in the variance of the average of all errors in the pixel (not just their sum), which we call $\sigma_P^2$:

\begin{align}
\sigma_P &= \frac{\sigma_{sum}}{P} \\
         &= \frac{\sqrt{P}RMS}{P} \\
         &= \frac{RMS}{\sqrt{P}}
\end{align}

The next task is to estimate P, the number of hits per pixel coming from independent scans. The LD flight hit map is shown in figure \ref{fig: hitmap} for pixels of size $\sim$6.9'
(nside=516 in the HEALPix pixelization scheme).  For pixels of size $\sim$1.7' (nside=2048), we can extrapolate from this figure that most pixels have between 500 and 7000 hits,
with 1600 hits on average. We do not want to count hits within a given throw because consecutive pointing errors are extremely correlated and do not average down. Because of our
scan patterns, we assume that a pixel is either hit consecutively within a single throw, or from different revisits from independent scans. Let's estimate the number of consecutive
hits of a pixel within a throw. The median velocity of the telescope is 0.1 \degree/s, and so it takes
$\sim$0.28 seconds to traverse the 1.7' pixel. The detectors are sampled at 191 Hz, and so they hit the pixel 53 times during the 0.28 seconds. So we estimate P by making the
approximation that all pixels have the same number of hits (1600) and then calculate that the number of hits with uncorrelated pointing errors is P = 1600/53 $\sim$30.  Since for
EBEX the requirements are that $\sigma_P \sim$10", the requirements on the RMS over a throw are RMS = $\sigma_P \sqrt{P} \sim $54".

\section{Conventions used for orthogonalization matrix O}
\label{appendix: rot and misal}

\begin{figure}
\centering
\includegraphics[scale=0.5]{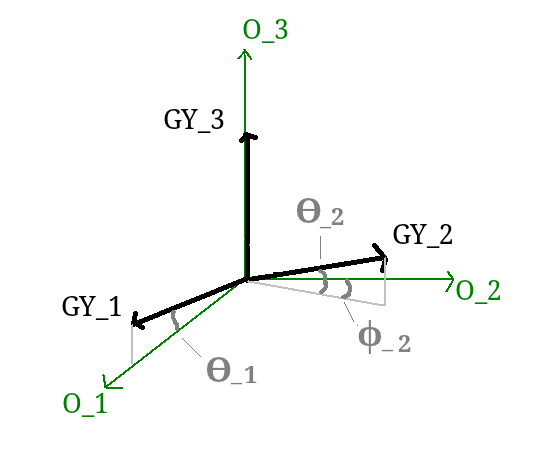}
\caption{Conventions for gyroscopes misalignment.}
\label{fig: gyro orthog}
\end{figure}

The gyroscopes are mounted close to an orthogonal basis labelled $(O_1, O_2, O_3)$, as shown in figure \ref{fig: gyro orthog}. One of the gyroscopes can be aligned with an axis of the
orthogonal basis without loss of generality.

At a given time step, the telescope rotates with a rotation vector $\vec{\omega}$. Because $(O_1, O_2, O_3)$ is an orthogonal basis, $\vec{\omega}$ can be expressed as:

\begin{equation}
\vec{\omega} = \omega_{O_1} \hat{o}_1 + \omega_{O_2} \hat{o}_2 + \omega_{O_3} \hat{o}_3
\end{equation}

The rotation measured by each gyroscope can be similarly expressed:

\begin{equation}
\left( \begin{array}{c}
\omega_{1} \\ \omega_{2} \\ \omega_{3} 
\end{array} \right)
= \left( \begin{array}{ccc}
\hat{o}_1 \cdot \hat{g}_1 &  \hat{o}_2 \cdot \hat{g}_1 & \hat{o}_3 \cdot \hat{g}_1 \\
\hat{o}_1 \cdot \hat{g}_2 &  \hat{o}_2 \cdot \hat{g}_2 & \hat{o}_3 \cdot \hat{g}_2 \\
\hat{o}_1 \cdot \hat{g}_3 &  \hat{o}_2 \cdot \hat{g}_3 & \hat{o}_3 \cdot \hat{g}_3 
\end{array} \right)
\left( \begin{array}{c}
\omega_{O_1} \\ \omega_{O_2} \\ \omega_{O_3} 
\end{array} \right)
\end{equation}

\noindent which can be re-written as:

\begin{equation}
\label{eq: O inverse}
\left( \begin{array}{c}
\omega_{1} \\ \omega_{2} \\ \omega_{3} 
\end{array} \right)
= \left( \begin{array}{ccc}
c_{\theta_1}   & 0 &  s_{\theta_1}\\
c_{\theta_2} s_{\phi_2}  & c_{\theta_2}c_{\phi_2} & s_{\theta_2} \\
0 &  0  & 1                           
\end{array} \right)
\left( \begin{array}{c}
\omega_{O_1} \\ \omega_{O_2} \\ \omega_{O_3} 
\end{array} \right)
\end{equation}

where all angles defined are small, of order arcminute. The UKF uses the matrix O, defined as the inverse of the matrix from equation \ref{eq: O inverse}.

\end{document}